\documentclass[conference]{IEEEtran}
\IEEEoverridecommandlockouts
\usepackage{cite}
\usepackage{amsmath,amssymb,amsfonts}
\usepackage{algorithmic}
\usepackage{graphicx}
\usepackage{textcomp}
\usepackage{xcolor}
\usepackage{subfig}
\usepackage{ntheorem}
\usepackage{url}
\usepackage{tikz}
\usepackage{booktabs}

\makeatletter
\newcommand{\figcaption}[1]{\def\@captype{figure}\caption{#1}}
\newcommand{\tblcaption}[1]{\def\@captype{table}\caption{#1}}
\makeatother

\newboolean{showcomments}
\setboolean{showcomments}{true} 
\ifthenelse{\boolean{showcomments}}
{\newcommand{\nb}[2]{
		\fcolorbox{black}{yellow}{\bfseries\sffamily\scriptsize#1}
		{\sf\small$\blacktriangleright$\textit{#2}$\blacktriangleleft$}
	}
	
}
{\newcommand{\nb}[2]{}
	
}

\usepackage[most]{tcolorbox}
\newtcolorbox{exmaplebox}[1]{
    colback=gray!5!white,
    colframe=gray!50!black,
    left=5pt, right=5pt, top=3pt, bottom=3pt,
    arc=2pt, boxrule=0.5pt,
    fonttitle=\bfseries,
    fontupper=\footnotesize,
    before upper={\setlength{\baselineskip}{11pt}},
    title=#1
}

\def\BibTeX{{\rm B\kern-.05em{\sc i\kern-.025em b}\kern-.08em
    T\kern-.1667em\lower.7ex\hbox{E}\kern-.125emX}}
\begin{document}
\bstctlcite{IEEEexample:BSTcontrol}

\title{
Bridging the Interpretation Gap in Accessibility Testing: Empathetic and Legal-Aware Bug Report Generation via Large Language Models
}

\author{
    \IEEEauthorblockN{Ryoya Koyama}
    \IEEEauthorblockA{\textit{Institute of Science Tokyo}\\
    Tokyo, Japan\\
    koyama.r.8474@m.isct.ac.jp}

    \\

    \IEEEauthorblockN{Jialong Li*}\thanks{* Corresponding Author: lijialong@fuji.waseda.jp}
    \IEEEauthorblockA{\textit{Waseda University}\\
    Tokyo, Japan\\ 
    lijialong@fuji.waseda.jp}
    
    \and
    \IEEEauthorblockN{Zhiyao Wang}
    \IEEEauthorblockA{\textit{The University of Osaka}\\
    Osaka, Japan\\
    wangzhiyao@ist.osaka-u.ac.jp}
    \\


    \and

    \IEEEauthorblockN{Devi Karolita}
    \IEEEauthorblockA{\textit{Palangka Raya University}\\
    Palangkaraya, Indonesia\\
    devikarolita@it.upr.ac.id}
    
    \\
    \IEEEauthorblockN{Kenji Tei}
    \IEEEauthorblockA{\textit{Institute of Science Tokyo}\\
    Tokyo, Japan\\
    tei@comp.isct.ac.jp}

}

\maketitle
\let\thefootnote\relax

\begin{abstract}
Modern automated accessibility testing tools for mobile applications have significantly improved the detection of interface violations, yet their impact on remediation remains limited. A key reason is that existing tools typically produce low-level, technical outputs that are difficult for non-specialist stakeholders, such as product managers and designers, to interpret in terms of real user harm and compliance risk. In this paper, we present \textsc{HEAR} (\underline{H}uman-c\underline{E}ntered \underline{A}ccessibility \underline{R}eporting), a framework that bridges this interpretation gap by transforming raw accessibility bug reports into empathetic, stakeholder-oriented narratives. Given the outputs of the existing accessibility testing tool, \textsc{HEAR} first reconstructs the UI context through semantic slicing and visual grounding, then dynamically injects disability-oriented personas matched to each violation type, and finally performs multi-layer reasoning to explain the physical barrier, functional blockage, and relevant legal or compliance concerns. We evaluate the framework on real-world accessibility issues collected from four popular Android applications and conduct a user study (N=12). The results show that \textsc{HEAR} generates factually grounded reports and substantially improves perceived empathy, urgency, persuasiveness, and awareness of legal risk compared with raw technical logs, while imposing little additional cognitive burden. \\
Reproducibility Package: \url{https://github.com/HaRRrrusame/hear-a11y-llm}.
\end{abstract}



\section{Introduction}

Mobile applications have become the primary gateway for essential daily activities, ranging from financial banking and healthcare management to social interaction and emergency services. Ensuring equitable access to these digital ecosystems is not merely a legal obligation mandated by regulations such as the Americans with Disabilities Act (ADA) and the European Accessibility Act (EAA), but a fundamental human right. According to the World Health Organization, over 1 billion people---approximately 16\% of the global population---live with some form of disability \cite{who_disability}. Consequently, the software engineering community has studied automated techniques to detect accessibility barriers in mobile apps.

In recent years, the landscape of automated accessibility testing has evolved significantly. Industry-standard tools like Google Accessibility Scanner \cite{google_scanner} and Lint-based static analyzers provide developers with immediate feedback on interface compliance. More recently, the research community has introduced dynamic crawling frameworks, such as \textit{Groundhog} \cite{salehnamadi2022groundhog} and \textit{Latte} \cite{salehnamadi2021latte}, which simulate user interactions to uncover deep-seated issues that manifest only during runtime execution. These tools have successfully lowered the technical barrier for \textit{detection}, enabling the identification of thousands of violations across millions of apps.

However, a critical issue remains: while the capability to \textit{detect} accessibility bugs has improved, the rate of \textit{remediation} (fixing) remains disappointingly low. A key reason for this stagnation is not the absence of data, but the \textit{inaccessibility of the data itself} to non-technical stakeholders. Current tools typically output low-level, syntax-oriented logs---such as JSON files containing view hierarchies, coordinate bounds, and cryptic error codes (e.g., \textit{bounds=[10,10][20,20], expected 48dp}).

This technical representation creates a profound "interpretation gap" within software development teams. Product Managers (PMs) and UI Designers, who typically control the development roadmap and resource allocation, often lack the specialized knowledge to map these technical metrics to actual user experiences. To a PM focused on conversion rates, a log entry stating "Touch target size is 32dp" may appear as a trivial "polish" issue with low priority. They fail to perceive the semantic reality: that for a user with Parkinson's disease, this "minor" issue effectively blocks a critical payment workflow. Consequently, accessibility fixes are frequently deprioritized in favor of new features, simply because the disability user impact and legal risk of the barriers are lost in interpretation.

To this end, this paper proposes \textsc{HEAR} (\underline{H}uman-c\underline{E}ntered \underline{A}ccessibility \underline{R}eporting), an automated framework designed to bridge this interpretation gap by transforming cold, code-centric testing results into persuasive, human-centric narratives that evoke stakeholder empathy and highlight urgency.
Specifically, \textsc{HEAR} primarily consists of two key steps. First, it dynamically maps accessibility violations to specialized user personas; for instance, a "small touch target" violation is mapped to the specific lived experience of a user with Parkinson’s disease who struggles with motor precision. Second, the framework enriches these narratives by integrating dual-perspective impact assessments: it simulates the user-centric hurdle to demonstrate how a bug leads to workflow abandonment, while simultaneously providing a regulatory-centric risk analysis based on local accessibility laws. By synthesizing these perspectives into an intuitive, high-level report, our framework empowers stakeholders to recognize the functional severity and legal vulnerability underlying each accessibility bug.

To summarize, this paper makes the following contributions:
\begin{itemize}
    \item We identify the challenge of the ``interpretation gap'' in current accessibility testing workflows, highlighting how the lack of semantic context in technical logs hinders effective remediation by stakeholders.
    \item We introduce \textsc{HEAR}, an automated framework that reconstructs UI context from decoupled logs and screenshots. By dynamically injecting ability-based personas and utilizing Chain-of-Thought (CoT) reasoning, the framework translates technical violations into human-centric narratives of user struggle and legal risk.
    \item We evaluate the reliability of \textsc{HEAR} on 103 bug instances (from four popular Android applications and conduct a user study ($N=12$) to assess its utility. 
\end{itemize}

The remainder of this paper is organized as follows. Section~\ref{sec:background} provides the necessary background. Section~\ref{sec:approach} details our proposed framework. Section~\ref{sec:evaluation} presents the experimental setup and results. Section~\ref{sec:Related} discusses related work. Finally, Section~VI concludes the paper and outlines future directions.

\section{Background}
\label{sec:background}

\subsection{Automated Accessibility Testing}
In the field of modern mobile accessibility testing, Google's ecosystem provides the de facto standard for the Android platform. The core of this ecosystem is the Accessibility Testing Framework (ATF), a library that performs static analysis on Android UI components. Tools such as the Google Accessibility Scanner \cite{} and Android Lint utilize ATF to traverse the View Hierarchy ($T_{xml}$) of a rendered screen, evaluating individual UI elements against a set of heuristic rules derived from the Web Content Accessibility Guidelines (WCAG).

These automated tools typically prioritize three primary categories of verification. First, they validate Touch Target Size, ensuring that interactive elements possess a density-independent pixel (dp) size of at least $48 \times 48$dp to accommodate users with limited fine motor control. Second, they perform Content Labeling checks, verifying that visual elements, such as an \texttt{ImageView}, possess textual descriptions (e.g., \texttt{contentDescription}) necessary for screen readers to convey meaning to visually impaired users. Third, they analyze Color Contrast, calculating the ratio between foreground text and background colors to ensure readability for users with low vision. Despite their scalability, the output of these tools is inherently low-level and data-centric. Results are typically serialized as JSON containing geometric bounds and cryptic error codes, lacking the semantic context required for developers to fully understand the user's struggle.

\subsection{Personas in Software Engineering}
In the domain of Software Engineering (SE), personas have emerged as a critical technique for requirements engineering and user-centered design \cite{KAROLITA2023107264}. A persona is a fictional, yet realistic, archetype representing a specific segment of users. In SE workflows, personas serve as a communication tool to facilitate a shared understanding of user needs among diverse stakeholders, including product managers, designers, and developers. They are instrumental in preventing the "Elastic User" phenomenon, where engineers unconsciously design software for themselves rather than for the intended audience.

Within the specific context of inclusive software development, researchers advocate for ability-based personas. Unlike traditional marketing personas that focus on demographics, ability-based personas model the functional constraints and assistive technologies of users, such as a user with "low peripheral vision relying on 200\% dynamic text resizing." However, a significant challenge in current SE practice is that personas are often treated as static artifacts, such as PDF documents or posters—that remain disconnected from the active testing and debugging workflow. Consequently, the empathetic connection is often lost by the time a developer addresses a specific bug report. 

\section{Proposal}
\label{sec:approach}

\subsection{Overview}
\begin{figure*}[ht!bp] 
    \centering
    \includegraphics[width=\textwidth]{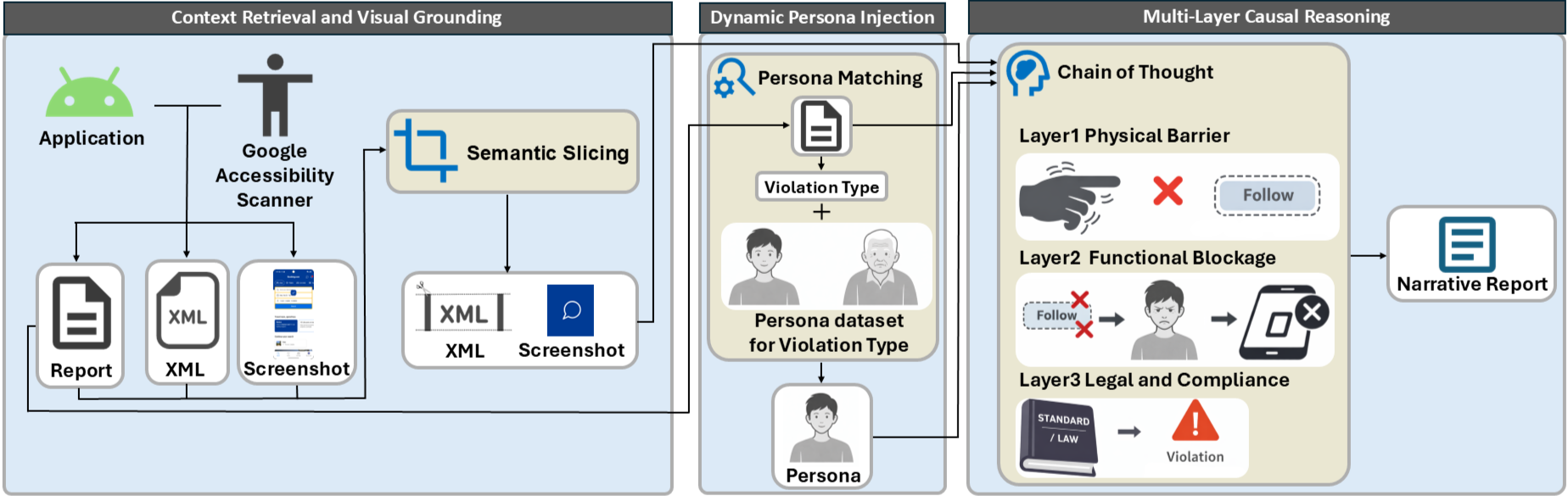}
    \caption{Overview of \textsc{HEAR}.} 
    \label{fig:architecture}
\end{figure*}

The core intuition behind \textsc{HEAR} is that an accessibility violation is not merely a syntactic error (e.g., \textit{bounds too small}), but a rupture in the user's interaction journey. As illustrated in Figure \ref{fig:architecture}, our pipeline takes raw bug reports and UI states as input, and outputs tailored narratives through a three-phase process: (i) \textbf{context retrieval \& visual grounding} to reconstruct the semantic scene from decoupled logs and screenshots; (ii) \textbf{dynamic persona injection} to map technical error types to specific user impairments (e.g., Parkinson's disease); and (iii) \textbf{multi-Layer causal reasoning} to utilize CoT to deduce physical barriers, functional blockages, and legal/compliance concerns.

\subsection{Phase I: Context Retrieval and Visual Grounding}
Existing accessibility testing tools often report violations in isolated JSON formats, which lack the visual and structural context necessary for high-level reasoning. To address this, our framework first restores the environment of each detected bug through a context retrieval process. By utilizing the unique coordinates and state identifiers of each bug instance, the system aligns the technical report with its corresponding UI artifacts—specifically, the View Hierarchy tree and raw high-resolution screenshots.

To ensure efficient processing by LLMs, we implemented a ``semantic slicing" mechanism to manage the high dimensionality of UI data. Rather than processing the entire XML tree, which introduces significant noise and token overhead, the slicer extracts a localized semantic window consisting of the target element’s attributes and its child elements.

This textual context is augmented by a visual grounding step, where it crops the original screenshot around the bug’s coordinates with 20\% padding. This padding provides the model with essential peripheral information regarding the element’s surroundings. The resulting dual-modal input enables the model to simultaneously analyze an element’s visual appearance and its structural role within the interface, providing a grounded foundation for accurate interaction simulation.

\begin{exmaplebox}{Example: Output of Google Accessibility Scanner}  
    Violation: Touch target \\
    Key/Bounds: [571,1952][795,2064] \\
    Description: Consider making this clickable item larger. This item's height is 43dp. Consider making the height of this touch target 48dp or larger.
\end{exmaplebox}

\subsection{Phase II: Dynamic Persona Injection}
A static explanation of accessibility rules fails to evoke empathy. To address this, we propose a Dynamic Persona Injection mechanism that selects a specific, detailed user profile based on the nature of the detected barrier. We formalize a persona $P$ as a set of constraints:$$P = \langle \text{Name}, \text{Loc}, \text{Cond}, \text{Cons}, \text{Psy}, \text{Log} \rangle$$where Name represents the name and age, Loc is the geographic jurisdiction, Cond is the medical condition, Cons represents the resulting physical constraints, Psy indicates psychological traits, and Log describes the specific logical difficulties encountered.The medical conditions for each persona are selected based on the Success Criteria defined by the Web Content Accessibility Guidelines (WCAG) corresponding to each accessibility violation. To enable the LLM to generate narratives that evoke realistic empathy, we define not only the medical conditions but also the subsequent physical constraints, psychological characteristics, and the specific difficulties anticipated within the context of application usage.

\begin{exmaplebox}{Example: Persona}  
     \textbf{Name: Ichiro (35)}
     \begin{itemize}
        \item \textit{Location:} Japan.
        \item \textit{Condition:} Cerebral Palsy - Athetoid type.
        \item \textit{Constraints:} Tap position deviates by approximately $\pm 40px$ due to involuntary movements.
        \item \textit{Psychology:} Low frustration threshold for operational errors.
        \item \textit{Logic:} Due to involuntary movements, taps land outside the boundaries of buttons.
     \end{itemize}
\end{exmaplebox}

\subsection{Phase III: Multi-Layer Causal Reasoning (CoT)}

We employ CoT prompting to bridge the gap between "low-level geometry" and "high-level legal risk." CoT is applied here because accessibility barriers are inherently \textit{context-aware}: the severity of a violation is determined by the interplay between the UI's functional role (e.g., a "Submit" button in a banking app) and the specific physical limitations of the user. Without explicit causal steps, LLMs tend to produce generic descriptions that lack the logical depth required to justify legal non-compliance or business urgency.

We structure the reasoning process into three layers:

\textit{Layer 1: Physical Barrier}.
The framework begins by leveraging the LLM's reasoning capabilities to simulate the direct interaction between the persona's specific constraints and the identified UI elements. Specifically, \textsc{HEAR} injects the full persona profile into the model's context, prompting it to simulate a realistic interaction attempt. By requiring the model to "act" as the defined persona, it can deduce the spatial struggle that occurs when the user’s physical limitations intersect with the element's bounding box and attributes retrieved in Phase I.

\begin{exmaplebox}{Example: Output of Layer 1}  
    Ichiro, a 35-year-old user from Japan with athetoid cerebral palsy, experiences involuntary muscle spasms that cause touch deviations of approximately ±40px. The current UI element, a "Follow" button, has a touch target height of 43dp, which is smaller than the recommended 48dp. This size discrepancy makes it difficult for Ichiro to accurately interact with the button, as his touch may frequently land outside the defined boundaries due to his condition.
\end{exmaplebox}

\textit{Layer 2: Functional Blockage}.
In the second layer, the model transitions from physical struggle to operational failure by analyzing the semantic role of the target element. By parsing the \texttt{resource-id}, neighboring text labels, and structural context within the View Hierarchy, \textsc{HEAR} deduces how the physical barrier from the previous step disrupts the user's intended workflow. This translation aims to shift the focus from a "localized bug" to a "service failure." Demonstrating the ripple effect, from a failed tap to the total abandonment of a critical task, provides stakeholders with a clear understanding of the bug's functional impact on user retention.

\begin{exmaplebox}{Example: Output of Layer 2}  
    As a result of the inadequate touch target size, Ichiro may struggle to successfully activate the "Follow" button. This can lead to repeated failed attempts, causing frustration and potentially discouraging him from using the application altogether. The operational error not only hinders his ability to follow other users but also impacts his overall user experience negatively.
\end{exmaplebox}

\begin{exmaplebox}{Example: Output of Layer 3}  
    - JIS X 8341-3:2016 Violation:\\
    \quad - Success Criterion Violated: 2.5.5 Target Size\\
    \quad - Requirement: The standard requires that touch targets should be at least 48dp by 48dp to ensure accessibility for users with motor impairments. This provision is intended to accommodate users who may have difficulty with precise touch interactions.\\
    - Legal Risks and Compliance Concerns:\\
    \quad - Prohibition of Unjust Discriminatory Treatment: By not adhering to accessibility standards, the application risks being perceived as engaging in unjust discriminatory treatment against users with disabilities, violating the Act on the Elimination of Discrimination against Persons with Disabilities in Japan.
\end{exmaplebox}

\textit{Layer 3: Legal and Compliance Concerns}.
Finally, the framework evaluates the functional blockage against specific regional mandates through a context-aware verification process. Utilizing the $\mathbf{Loc}$ (Location) attribute from the Persona, the system dynamically retrieves relevant legal clauses and technical standards from a localized regulatory database—such as specific criteria from JIS X 8341-3 for Japan or articles from the EAA for Europe. These retrieved legal texts are then injected into the model's prompt as grounded context, enabling the LLM to perform a direct comparative analysis between the identified functional blockage and the statutory requirements. By explicitly citing these retrieved laws, the system transforms a subjective usability concern into a documented "legal vulnerability," providing stakeholders with the objective justification needed for immediate remediation.

\section{Evaluation}
\label{sec:evaluation}

The evaluation aims to address the following two RQs:

\begin{itemize}
    \item \textbf{RQ1 (Accuracy \& Reliability):} To what extent can \textsc{HEAR} generate factually accurate narratives without hallucinating UI elements or functional consequences?
    \item \textbf{RQ2 (Utility \& Persuasiveness):} How does the persona-based narrative influence stakeholders' perception of bug severity and prioritization compared to raw logs?
\end{itemize}

\subsection{Experimental Setup}

\textit{Target Application}.
We selected four representative apps across the Social, Shopping, Business, and Travel categories: \textit{Instagram, Wish, Teams, and Booking}, as targets for extracting accessibility violations. This selection covers a diverse range of categories, providing a generalized foundation for evaluating our framework while maintaining a manageable scale for the user-study assessment.
We executed the Google Accessibility Scanner on four representative screens of each app and randomly extracted $N=103$ accessibility violations covering three common error categories. The breakdown is as follows: \texttt{TouchTargetSize} (61\%), \texttt{ContentLabeling} (13\%), and \texttt{ContrastRatio} (26\%). For each violation, we extracted the corresponding screen screenshot ($I_{scr}$), the view hierarchy ($T_{xml}$), and the raw bug report ($B_{raw}$) generated by the Google Accessibility Scanner.
For each type of violation, we developed two representative personas based on WCAG guidelines. Furthermore, to align with the cultural background of our user experiments, we designated Japan as the geographic jurisdiction. This led us to ground the legal rationales in JIS X 8341-3:2016 and the Act on the Elimination of Discrimination against Persons with Disabilities in our experiment.

\textit{Baselines}.
We compare our approach against the raw logs: The unmodified JSON output from Google Accessibility Scanner.

\textit{LLM Choice}.
For implementation, we utilize GPT-4o as the underlying LLM. The temperature is set to 0.1 to ensure deterministic output while allowing for natural phrasing.

\subsection{Experiment Settings for RQ1}
To quantify the reliability of the generated reports, we developed a structured annotation codebook. Two authors independently audit 103 random samples.

Each generated report is evaluated against the ground truth screenshot and the corresponding XML tree on a binary pass/fail scale across three distinct dimensions. (i) Visual grounding assesses whether the UI element described in the narrative, such as a specific icon or button, actually exists at the coordinates specified in the technical log. (ii) Textual fidelity verifies that any quoted strings, including labels, titles, or neighboring textual context, are verbatim matches to the information present in the screenshot and view hierarchy. (iii) Functional logic evaluates whether the inferred interaction consequence, such as a user being unable to complete a checkout process, is a sound and plausible deduction based on the overall UI context and functional purpose of the screen. A report is classified as a hallucination if it fails to meet the requirements of any single dimension.

\subsection{Experiment Settings for RQ2}
We conduct a within-subject controlled experiment to measure stakeholder perception across diverse development scenarios.
The experimental stimuli consist of a scenario featuring one of three types of accessibility bugs: small touch targets, unlabeled navigation icons, or low-contrast text elements.

Each scenario is presented to the participants in two distinct formats: (i) a baseline representing the raw JSON output from standard scanning tools, and (ii) a narrative report generated by our framework. To mitigate potential learning effects or fatigue, the presentation order of the scenarios and formats is randomized for each participant. Regarding the survey instrument, participants evaluate each report based on a 5-point Likert scale, ranging from strongly disagree to strongly agree, covering dimensions such as clarity, empathy, perceived urgency, and persuasiveness, as detailed in Table~\ref{tab:questionnaire}

\begin{table*}[h]
\centering
\caption{User Study Questionnaire Items (Per-Scenario)}
\label{tab:questionnaire}
\begin{tabular}{p{0.25\linewidth} p{0.65\linewidth}}
\toprule
\textbf{Dimension} & \textbf{Question Item} \\
\midrule
\textbf{Q1. Clarity} & "It is easy for me to visualize the exact problem described in this report." \\
\midrule
\textbf{Q2. Empathy} & "I can vividly imagine the frustration or difficulty experienced by the user with disabilities." \\
\midrule
\textbf{Q3. Urgency} & "Based \textit{solely} on this report, I would classify this bug as a 'Critical' priority that blocks release." \\
\midrule
\textbf{Q4. Persuasion} & "This report provides sufficient justification for me to convince stakeholders (e.g., clients/executives) to fix it." \\
\midrule
\textbf{Q5. Cognitive Load} & "I had to invest significant mental effort to understand the meaning of this bug." (Reverse Scale) \\
\midrule
\textbf{Q6. Legal/Compliance Risk} & "After reading this report, I am concerned about potential legal or compliance issues if this bug is not fixed." \\
\bottomrule
\end{tabular}
\end{table*}

Additionally, after reviewing all experimental scenarios, participants complete a final evaluation to assess the utility of the two reporting formats across four distinct development lifecycle tasks. Using a 5-point preference scale, where 1 indicates a definite preference for the baseline and 5 indicates a definite preference for our framework, participants evaluate the formats in the context of (i) locating the specific code segment and implementing the technical fix, (ii) explaining the issue to non-technical stakeholders such as product managers or clients, (iii) deciding the priority level of the bug fix within the development backlog, and (iv) learning about accessibility concepts and diverse user needs. For the task-based preferences, we analyze the distribution of participant choices to identify the specific strengths of each reporting format across different stages of the software development process.

We recruited $N = 12$ participants (11 male, 1 female) from the university campus. The participants comprised 4 PhD students, 6 Master's students, and 2 fourth-year undergraduates, with primary research areas in Software Engineering ($N = 7$), AI/Machine Learning ($N = 3$), and HCI/UI/UX ($N = 2$). In terms of programming experience, 7 reported 3 or more years, 1 reported 1 to 3 years, 2 reported less than 1 year, and 2 reported none. Mobile app development experience was limited, with 10 reporting none, 1 having class or hobby-level experience, and 1 having professional experience. For software testing and debugging, 5 had class-level experience, 4 had professional experience, 2 had none, and 1 reported it as a routine practice. Prior knowledge of software accessibility was generally low: 7 reported no knowledge of the concept, 2 had basic knowledge, 2 had learned knowledge through guidelines or coursework, and 1 had practical implementation experience. Regarding generative AI usage, 9 participants reported using tools such as ChatGPT or Gemini on an almost daily basis, and 3 reported using them several times per week.

\subsection{Experiment Results}

In this section, we present the quantitative and qualitative findings regarding the accuracy of our generated reports and their impact on stakeholder perception.

For RQ1, we conducted annotations on 103 bug scenarios collected from four real-world applications. The annotation codebook was established prior to the formal evaluation, and a pilot annotation was carried out to calibrate the criteria and improve consistency in applying the three coding dimensions. In the formal annotation phase, we did not observe hallucinations in the confirmed cases we examined across visual grounding, textual fidelity, and functional logic. No disagreements were observed in the independent scoring results. These findings suggest that a modern commercial LLM can accurately interpret the provided UI context and screenshots to produce reliable, fact-grounded narratives in our evaluation setting.

\begin{figure}[h!tbp]
    \centering
    \includegraphics[width=\dimexpr\linewidth-2\fboxrule-2\fboxsep\relax]{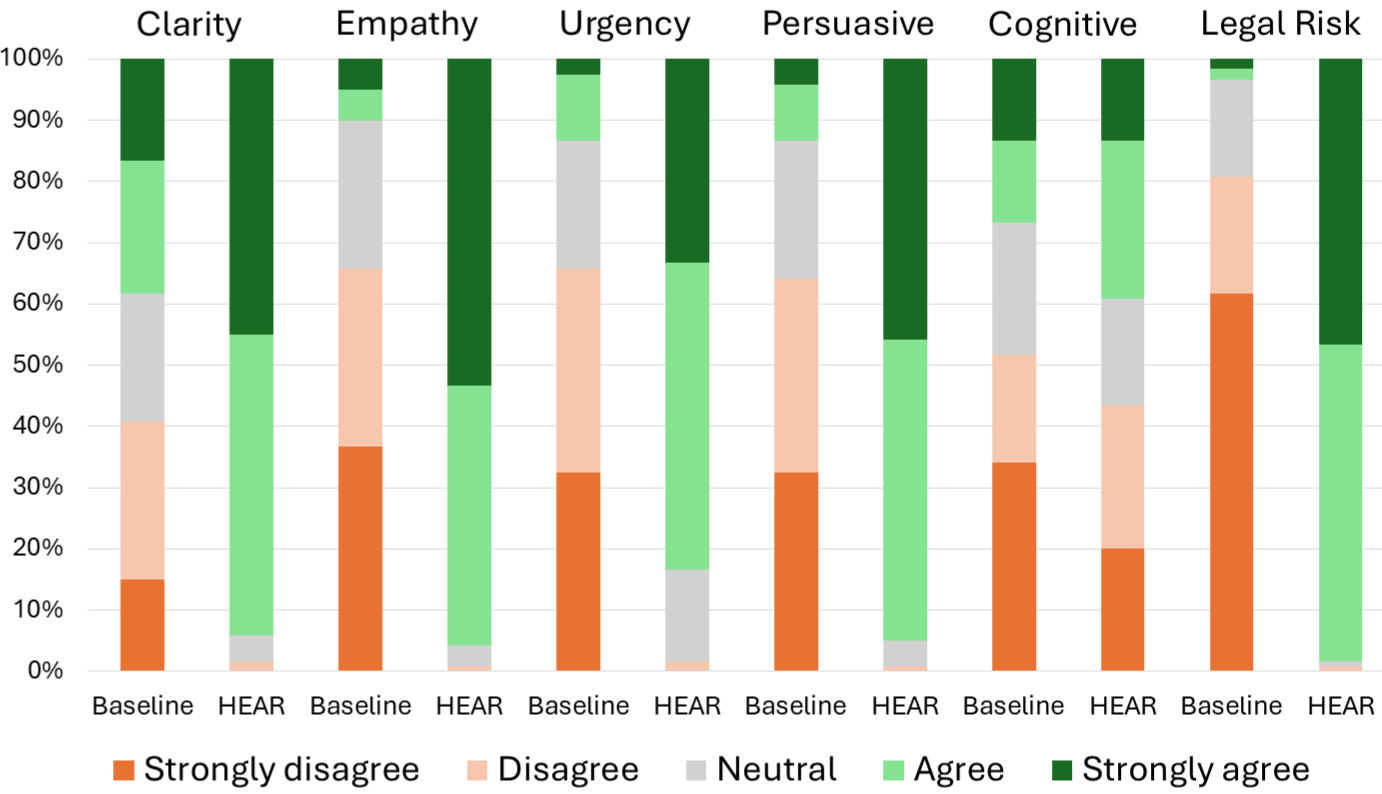}
    \caption{Stakeholder perception scores comparing Baseline (Raw JSON) vs. \textsc{HEAR}. }
    \label{fig:likert_comparison}
\end{figure}

For RQ2, Figure \ref{fig:likert_comparison} compares stakeholder perceptions of the baseline (Raw JSON) and \textsc{HEAR} across six evaluation dimensions. Overall, \textsc{HEAR} consistently received more favorable ratings than the baseline. The largest improvements were observed in Empathy and Legal Risk, where responses shifted from predominantly negative or neutral under the baseline to overwhelmingly positive under \textsc{HEAR}. Similar trends were found for urgency and persuasiveness, suggesting that \textsc{HEAR} helped stakeholders recognize accessibility issues not merely as technical defects, but as problems requiring timely attention and action. \textsc{HEAR} also achieved higher ratings in Clarity, indicating that participants found the reports easier to understand. Notably, this improvement did not come at the cost of substantially greater cognitive load: although responses were somewhat more distributed, the overall pattern suggests that the richer contextual framing of \textsc{HEAR} increased interpretability and impact without making the reports significantly more difficult to process.

\begin{figure}[h!tbp]
    \centering
    \includegraphics[width=0.99\dimexpr\linewidth-2\fboxrule-2\fboxsep\relax]{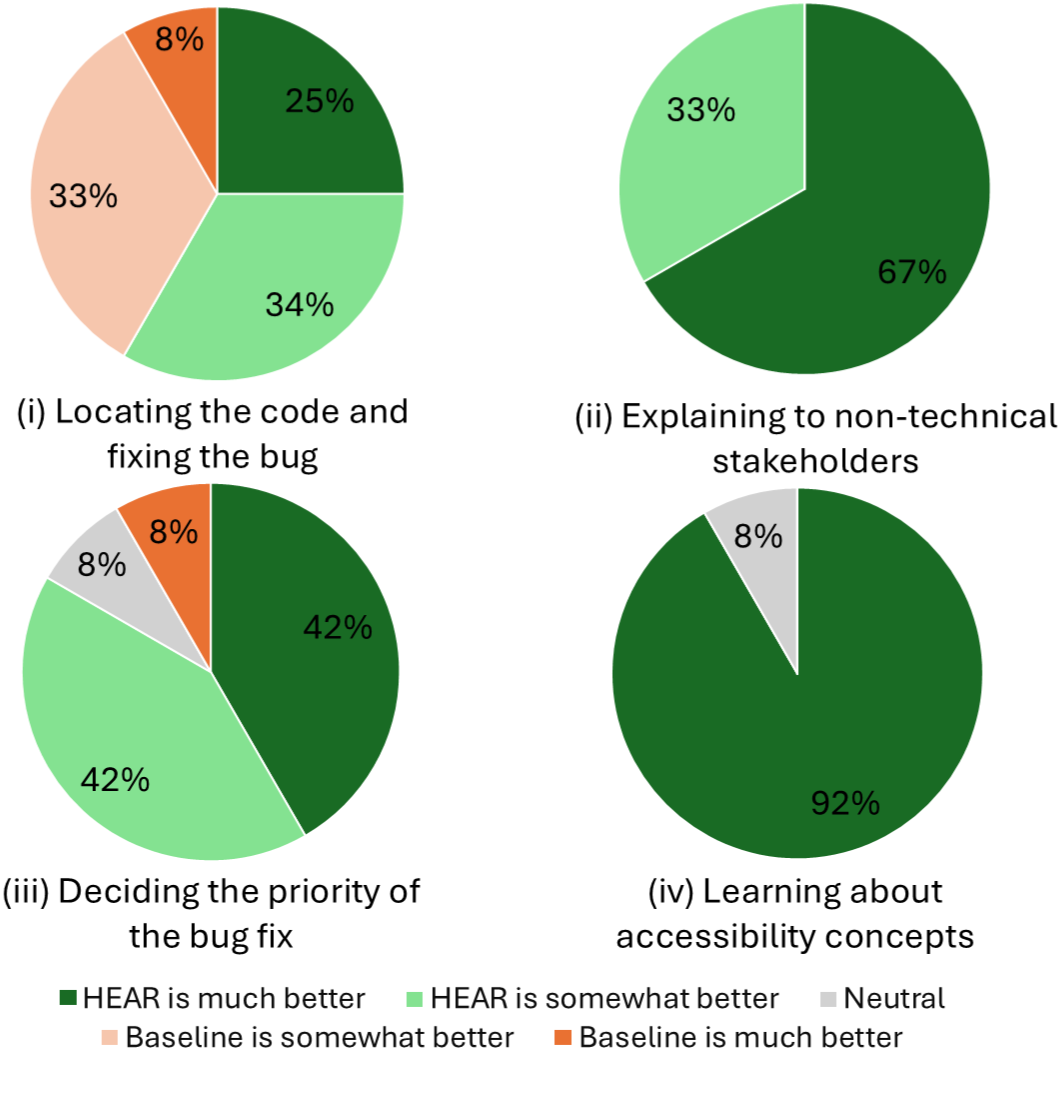}
    \caption{Participant preferences for report formats across four distinct development tasks.}
    \label{fig:task_preference}
\end{figure}

Figure \ref{fig:task_preference} shows that participants’ preferences varied by task, but consistently favored HEAR for communication- and understanding-oriented activities. The strongest preference emerged for learning about accessibility concepts, where 92\% of participants rated HEAR as much better and the remaining 8\% as neutral. A similarly strong pattern appeared for explaining issues to non-technical stakeholders, for which 100\% of participants preferred HEAR, including 67\% who considered it much better. HEAR was also clearly favored for deciding bug-fix priority, with 84\% of participants preferring it and only 8\% favoring the baseline. In contrast, preferences were more mixed for locating the relevant code and implementing the fix: although 59\% still favored HEAR, 41\% preferred the baseline. This suggests that HEAR is especially effective for improving conceptual understanding, stakeholder communication, and prioritization, whereas raw technical logs remain useful during hands-on debugging and implementation, where concise and low-level information is often more directly actionable.

\subsection{Discussion and Limitation}

Overall, the results suggest that \textsc{HEAR} helps bridge the gap between machine-detected accessibility violations and stakeholder understanding. It was particularly effective for communication-oriented tasks, improving perceived empathy, urgency, persuasiveness, and legal-risk awareness, and was strongly preferred for explaining issues to non-technical stakeholders, prioritizing fixes, and learning accessibility concepts. By contrast, preferences were more mixed for code-oriented debugging tasks, indicating that \textsc{HEAR} is better understood as a complementary reporting layer rather than a replacement for raw technical logs. These findings highlight the practical value of human-centered accessibility reporting while also revealing several limitations.

A primary limitation of \textsc{HEAR} is its inherent dependency on the fidelity of upstream detection tools. Our framework operates as a post-processing layer and fundamentally relies on the output of the underlying scanning tool (e.g., Google Accessibility Scanner). Consequently, the system inherits the false positive rates of these tools. If the scanner incorrectly flags a valid design as a violation, our LLM-driven approach may inadvertently generate a persuasive, empathetic narrative for a non-existent bug. This "amplification of error" risks validating false positives, potentially distracting developers from genuine issues and emphasizing the need for higher-precision detection mechanisms as a prerequisite for our pipeline.

Second, while the injection of dynamic personas effectively evokes empathy, it entails a necessary simplification of complex human conditions. Real-world disabilities are often intersectional and highly variable (e.g., a user managing both low vision and motor tremors), whereas our current model maps errors to specific, static profiles. This reductionism risks creating a "stereotype bias," where developers may unconsciously optimize software only for the specific traits of the generated persona while neglecting the broader, more nuanced spectrum of needs within that disability category.

\subsection{Threats to Validity}
\textit{Internal Validity.}
Threats to internal validity primarily stem from the subjectivity of manual assessments and potential confounding variables in the baseline comparison. Regarding RQ1, the classification of LLM output as "hallucination" or "logical inference" relies on human judgment. Although we employed two annotators to evaluate individually, this potentially introduced interpretative bias. Furthermore, regarding RQ2, comparing our persona-based narratives against raw JSON logs introduces a confounding variable of information density. Participants may prefer \textsc{HEAR} simply because it provides a clear natural language explanation, regardless of the specific "persona" element. It remains possible that the improved perception is driven by the textual clarity rather than the empathetic characterization, suggesting that the specific contribution of the persona injection may be conflated with the general benefits of summarization.

\textit{External Validity.}
External validity is limited by our participant demographics and dataset selection. We recruited computer science graduate students as proxies for industry practitioners; however, these participants lack the commercial pressures—such as strict release deadlines and revenue KPIs—that real-world product managers face. In a corporate environment, business constraints often override empathy, meaning the high "perceived urgency" reported by students may not fully translate to prioritization decisions in a high-pressure industrial setting. Additionally, our evaluation relied on four popular applications; these applications may not fully represent the diversity of modern UI paradigms in the millions of apps currently on the Google Play Store.

\textit{Construct Validity.}
A significant threat to construct validity lies in the potential gap between reported intent and actual behavior, often referred to as the "Say-Do Gap." Our survey instrument measured the perception of severity and the intent to prioritize bugs, rather than the actual action of fixing them. While our results indicate that \textsc{HEAR} successfully increases perceived urgency, the measured increase in empathy does not guarantee a corresponding increase in the actual remediation rate of accessibility barriers in a production environment.

\section{Related Work}
\label{sec:Related}

\subsection{Automated Accessibility Testing and the Reporting Gap}
The prevalence of accessibility barriers in mobile apps is well-documented, with recent studies showing that a vast majority of popular apps fail to meet basic inclusivity standards \cite{10.1145/3377811.3380392}. 
To address this, the community has evolved from static analysis tools \cite{10.1145/3218585.3218673}, 
which often suffer from high false positive rates, to dynamic, crawler-based frameworks. Notably, \textit{Latte} \cite{salehnamadi2021latte} and \textit{Groundhog} \cite{salehnamadi2022groundhog} introduced assistive-service driven testing, enabling the detection of deep-seated runtime issues that manifest only during interaction. 
However, while detection capabilities have advanced, the \textit{communication} of these findings remains stagnant. Existing reporting mechanisms typically output raw violations (e.g., view bounds, missing labels) \cite{10.1145/3551349.3560424}. Although \cite{10.1145/3674967} attempted to summarize accessibility reports to reduce redundancy, the output remains technical and data-centric. Our work addresses this "interpretation gap," arguing that the lack of semantic context in reports contributes to the low remediation rates observed in practice \cite{10.1145/3419249.3420119}.

\subsection{LLM-Driven UI Understanding and User Simulation}
The emergence of LLMs has provided new opportunities for semantic UI understanding. Models have demonstrated the ability to interpret mobile GUIs for automation tasks, as seen in \textit{DroidBot-GPT} \cite{wen2024droidbotgptgptpowereduiautomation} and conversational agents \cite{10.1145/3544548.3580895}. Specifically relevant to accessibility, \textit{HintDroid} \cite{10.1145/3613904.3642939} utilizes LLMs to predict missing content labels, while \cite{10.1145/3744257.3744270} explored LLMs for automated repair.
Beyond static understanding, recent advancements in user simulation, such as \textit{RecUserSim} \cite{10.1145/3701716.3715258} and other agent-based approaches \cite{10.1145/3708985, 10.1145/3696410.3714858,10773792, 10612015,10.1145/3643915.3644088}, utilize profile-memory-action architectures to mimic realistic user behaviors. These works demonstrate that LLMs can effectively adopt specific personas. However, these simulators primarily focus on recommender systems or general behavior, rather than simulating the physical and frustration dynamics of users with disabilities interacting with faulty interfaces.

\subsection{Empathy and Persona-Based Approaches}
Empathy is recognized as a cornerstone of inclusive design, essential for bridging the cognitive gap between developers and diverse user populations \cite{Kolko2015, 10.1145/3027063.3053347}. \cite{10.1145/3027063.3053347} highlighted that optimal user experience requires balancing "systemizing" (technical) and "empathizing" (human) aspects. Similarly, Maxim et al. \cite{10.1145/3486011.3486438} demonstrated the efficacy of empathic design in mobile contexts, while \cite{Postma2012} identified the significant industrial challenges in sustaining this empathy throughout the development lifecycle.
To operationalize empathy, the HCI community has extensively adopted \textit{Personas}. Beyond generic marketing profiles, researchers have advocated for \textit{ability-based personas} to specifically model disabilities. For instance, \cite{10.1145/3517428.3550364} developed personas for signing user populations based on ability ranges rather than medical deficits. \cite{app11010368} utilized persona cards to surface accessibility barriers for elderly users in home appliance interactions. 
However, creating these personas is traditionally a labor-intensive, manual process \cite{10.1145/3604571.3604572}. While recent work has explored generating synthetic interaction data for the elderly \cite{Maqbool1964352} or using empathy sandboxes for privacy education \cite{10.1145/3613904.3642363}, there remains a lack of automated mechanisms that can dynamically generate persona-driven narratives for specific software defects. \textsc{HEAR} fills this gap by automating the translation of technical logs into empathetic, persona-specific user stories.

\section{Conclusion and Future Work}
This paper addressed a key limitation of mobile accessibility testing: although current tools can detect violations effectively, their low-level outputs often fail to convey user impact to broader stakeholders. To bridge this interpretation gap, we proposed \textsc{HEAR}, a framework that converts raw scanner results into human-centered reports through UI context reconstruction, dynamic persona injection, and reasoning about functional and legal consequences.
Our evaluation of four real-world Android applications indicates that the generated reports remained grounded in the UI evidence and, compared with raw technical logs, improved participants’ understanding of user difficulty, urgency, persuasiveness, and compliance risk. At the same time, the user study indicates that it serves as a complementary reporting layer that supports prioritization and stakeholder communication, while technical logs remain useful for implementation and debugging.

In future work, we plan to extend our study in the following directions. First, we plan to enhance the granularity of our persona library by including intersectional disabilities and varying levels of digital literacy. This expansion will allow the system to model more complex user scenarios, such as an elderly user managing both low vision and cognitive decline. Second, we intend to integrate our framework with dynamic crawling tools like \textit{Groundhog} to move beyond static snapshot analysis. This will enable the LLM to ingest execution traces and narrate temporal navigation barriers that only manifest during runtime interaction. Third, we aim to conduct a longitudinal evaluation by deploying \textsc{HEAR} as a plugin within real-world CI/CD pipelines. By integrating with issue trackers like Jira, we can quantitatively measure the impact of humanized reports on actual bug-fix rates and development prioritization over time.

\section*{Acknowledgment}
This research was partially supported by JSPS KAKENHI (Grant Number 23K28064, 25K15290). The study protocol was granted an exemption by the Institutional Review Board (IRB) of the Institute of Science Tokyo. 
Prior to the start of the experiment, we obtained informed consent from each participant. Participants were fully briefed on the experimental procedures and were informed of their right to withdraw from the study at any time without penalty.

\bibliographystyle{IEEEtran}
\bibliography{bib}
\end{document}